\documentclass[prl,twocolumn,showpacs,superscriptaddress,preprintnumbers]{revtex4-2}
\usepackage{amsmath,amssymb,amsfonts,float,graphics,epsfig,epstopdf,color,verbatim,tabularx,bm,multirow,appendix}

\usepackage{graphicx}
\usepackage{xcolor}
\usepackage{dcolumn}
\usepackage{bm}
\usepackage{amsmath}
\usepackage{xspace}
\usepackage{time}

\pdfoutput=1
\usepackage{color}
\definecolor{LinkColor}{rgb}{0.256,0.439,0.588}
\usepackage{hyperref}
\hypersetup{
colorlinks=true,
citecolor=LinkColor,
linkcolor=LinkColor,
urlcolor=LinkColor
}

\newcommand{\beq}{\begin{equation}}
\newcommand{\eeq}{\end{equation}}
\newcommand{\beqn}{\begin{eqnarray}}
\newcommand{\eeqn}{\end{eqnarray}}

\newcommand{\ra}{\rightarrow}

\newcommand{\cA}{ {\cal A} }

\newcommand{\cH}{ {\cal H} }

\newcommand{\cL}{ {\cal L} }

\newcommand{\vect}[1]{{\bm{#1}}}

\newcommand{\ii}{\mathrm{i}}

\newcommand{\hn}{\hat{n}}

\newcommand{\SO}{\mathrm{SO}}

\newcommand{\U}{\mathrm{U}}

\newcommand\startsupplement{%
       \newpage\clearpage
       \setcounter{secnumdepth}{2}
       \setcounter{table}{0}
       \renewcommand{\thetable}{S\arabic{table}}
       \setcounter{figure}{0}
       \renewcommand{\thefigure}{S\arabic{figure}}
       \setcounter{equation}{0}
       \renewcommand{\theequation}{S\arabic{equation}}
       \setcounter{section}{0}
       \renewcommand{\thesection}{Section \Roman{section}}
       \renewcommand{\thesubsection}{\Roman{section}. \Alph{subsection}}
    }

\begin{document}

\title{Disorder Operator and R\'enyi Entanglement Entropy of Symmetric Mass Generation}

\author{Zi Hong Liu}
\affiliation{Institut für Theoretische Physik und Astrophysik and Würzburg-Dresden Cluster of Excellence ct.qmat,
	Universität Würzburg, 97074 Würzburg, Germany}

\author{Yuan Da Liao}
\affiliation{Department of Physics and HKU-UCAS Joint Institute
	of Theoretical and Computational Physics, The University of Hong Kong,
	Pokfulam Road, Hong Kong SAR, China}

\author{Gaopei Pan}
\affiliation{Department of Physics and HKU-UCAS Joint Institute
	of Theoretical and Computational Physics, The University of Hong Kong,
	Pokfulam Road, Hong Kong SAR, China}

\author{Menghan Song}
\affiliation{Department of Physics and HKU-UCAS Joint Institute
	of Theoretical and Computational Physics, The University of Hong Kong,
	Pokfulam Road, Hong Kong SAR, China}

\author{Jiarui Zhao}
\affiliation{Department of Physics and HKU-UCAS Joint Institute
	of Theoretical and Computational Physics, The University of Hong Kong,
	Pokfulam Road, Hong Kong SAR, China}

\author{Weilun Jiang}
\affiliation{State Key Laboratory of Quantum Optics and Quantum Optics Devices, Institute of Opto-Electronics, Shanxi University, Taiyuan 030006, China}

\author{Chao-Ming Jian}
\affiliation{Department of Physics, Cornell University, Ithaca, NY, USA}

\author{Yi-Zhuang You}
\affiliation{Department of Physics, University of California, San Diego, CA 92093, USA}

\author{Fakher F. Assaad}
\email{fakher.assaad@physik.uni-wuerzburg.de}
\affiliation{Institut für Theoretische Physik und Astrophysik and Würzburg-Dresden Cluster of Excellence ct.qmat,
	Universität Würzburg, 97074 Würzburg, Germany}

\author{Zi Yang Meng}
\email{zymeng@hku.hk}
\affiliation{Department of Physics and HKU-UCAS Joint Institute
	of Theoretical and Computational Physics, The University of Hong Kong,
	Pokfulam Road, Hong Kong SAR, China}
 
\author{Cenke Xu}
\email{xucenke@ucsb.edu}
\affiliation{Department of Physics, University of California, Santa Barbara, CA 93106}

\date{\today}

\begin{abstract}

The ``symmetric mass generation" (SMG) quantum phase transition discovered in recent years~\cite{slagleExotic2015,ayyarMassive2015,Catterall1510.04153,ayyarOrigin2016,Catterall1609.08541,he2016quantum,Kikukawa1710.11101,Kikukawa1710.11618,you2018symmetric,you2018bosonic,Xu2103.15865,Tong2022SMG,zeng2022symmetric,Wang1307.7480,YZY2022nuSMG,YZY2022FSSMG,wang2022symmetric}  has attracted great interests from both condensed matter and high energy theory communities. Here, interacting Dirac fermions acquire a gap  without condensing any fermion bilinear mass term or any concomitant spontaneous symmetry breaking. It is hence beyond the conventional Gross-Neveu-Yukawa-Higgs paradigm. One important question we address in this work is whether the SMG transition corresponds to a true unitary conformal field theory (CFT). We employ the sharp diagnosis including the scaling of disorder operator and R\'enyi entanglement entropy in large-scale lattice model quantum Monte Carlo simulations. Our results strongly suggest that the SMG transition is indeed an unconventional quantum phase transition and it should correspond to a true $(2+1)d$ unitary CFT.

\end{abstract}

\maketitle

\noindent{\textcolor{blue}{\it Introduction and Motivation.}---} When a quantum critical point (continuous quantum phase transition) does not have a Landau-Ginzburg type of description in terms of a local order parameter, it is often referred to as an unconventional QCP. In this work we carefully investigate a class of candidate unconventional QCPs involving Dirac fermions, which were proposed to be beyond the ``traditional" paradigm. When we discuss QCPs involving Dirac fermions, the standard paradigm is the Gross-Neveu-Yukawa-Higgs mechanism, in which a bosonic field would couple with the Dirac mass operator, and when the bosonic field condenses, the Dirac fermions acquires a mass~\cite{liaoDiracI2022,liu2022metallic,lang2019quantum,liu2021grossneveu}. The bosonic field carries certain representation of a symmetry group (or gauge group), hence the bosonic field plays the same role as the order parameter in the Landau-Ginzburg theory. In this conventional paradigm, the Dirac fermions acquire a mass through spontaneously breaking certain symmetry. This is essentially the mechanism for the mass generation of all the matter fields in the Standard Model of particle physics. However, in recent years it was discovered by both condensed matter and high energy physics communities that, under the right conditions, the Dirac fermions can acquire a mass continuously through a QCP {\it without} breaking any symmetry. This mechanism is called the ``symmetric mass generation" (SMG)~\cite{slagleExotic2015,ayyarMassive2015,Catterall1510.04153,ayyarOrigin2016,Catterall1609.08541,he2016quantum,Kikukawa1710.11101,Kikukawa1710.11618,you2018symmetric,you2018bosonic,Xu2103.15865,Tong2022SMG,zeng2022symmetric,Wang1307.7480,YZY2022nuSMG,YZY2022FSSMG,wang2022symmetric}. The possibility of SMG in $(2+1)d$ is tightly related to the classification of interacting topological insulators or topological superconductors in three spatial dimensions, please refer to Ref.~\cite{slagleExotic2015} for more discussions.

Another archetypal unconventional QCP is the ``deconfined quantum critical point" (DQCP) between the N\'eel and valence bond solid (VBS) orders, supposedly realized in certain frustrated spin-1/2 quantum magnets on the square lattice~\cite{senthil2004deconfined,senthil2004quantum}. In the last two decades, the nature of the DQCP has been studied with enormous efforts analytically, numerically and experimentally~\cite{senthil2004deconfined,senthil2004quantum,senthil2023deconfined,levin2004deconfined,senthil2005deconfined,senthil2006competing,sandvik2007evidence,melko2008scaling,lou2009antiferromagnetic,banerjee2010impurity,sandvik2010continuous,xu2012unconventional,pujari2013neel,block2013fate,harada2013possibility,levin2004deconfined,nahum2015deconfined,nahum2015emergent,shao2016quantum,wang2017deconfined,qin2017duality,d2017new,zhang2018continuous,maDynamical2018,huang2019emergent,roberts2019deconfined,ma2019role,xu2019monte,li2019deconfined,nahumNote2020,maTheory2020,sreejithEmergent2019,zhaoMulticritical2020,sandvik2020consistent,liaoDiracI2022,liu2022metallic,liaoTeaching2023,zhou2023mathrm,xi2023plaquette,chen2023phases,song2023deconfined,zayed4spin2017,guoQuantum2020,sunEmergent2021,cuiProximate2023,songUnconventional2023,Liu18,WangZ20,SatoT22,guoDeconfined2023}. Great progresses have been made regarding its potential emergent SO(5) symmetries~\cite{senthil2006competing,nahum2015deconfined,nahum2015emergent,ma2019role}, the surrounding duality web~\cite{wang2017deconfined,qin2017duality}, and the connection to the symmetry protected topological phase in the higher dimension~\cite{ashvinsenthil}, etc.. However, despite all these progress, the very nature of the DQCP, i.e. whether it corresponds to a true $(2+1)d$ unitary CFT or not, remains controversial. One indication that the DQCP should not be a true CFT is that, it ``failed" a series of general standards that all $(2+1)d$ unitary CFTs are expected to meet. These standards include the universal logarithmic contribution to both the disorder operator and the entanglement entropy defined in a subregion of the $2d$ space, when the subregion involves sharp corners~\cite{zhao2022scaling,wang2022scaling,liu2023fermion,liaoTeaching2023,song2023deconfined}~\footnote{A summary of these results and more analysis will be presented in an upcoming work.}.


Compared with the DQCP, the nature of SMG is even more difficult to address employing analytical techniques such as field theory due to the lack of a controlled limit (for example a generalization to a controlled large-$N$ limit), though a candidate field theory has been proposed in Refs.~\cite{you2018symmetric,you2018bosonic}. Hence we need to resort to numerical techniques. The previous quantum Monte Carlo (QMC) simulations suggest that the SMG could indeed be a continuous phase transition, i.e. the QCP of SMG indeed appears to be a $(2+1)d$ CFT~\cite{slagleExotic2015,ayyarMassive2015,Catterall1510.04153,ayyarOrigin2016,he2016quantum}, as the computed Dirac fermion mass increases continuously from zero at a critical strength of interaction. But the nature of SMG still needs to be tested using the same standards as DQCP on general grounds. In this work, we will employ the ``tests" that the DQCP failed to pass: the scaling of disorder operator and the R\'enyi entanglement entropy from lattice model QMC simulations.

\begin{figure}[htp!]
\includegraphics[width=\columnwidth]{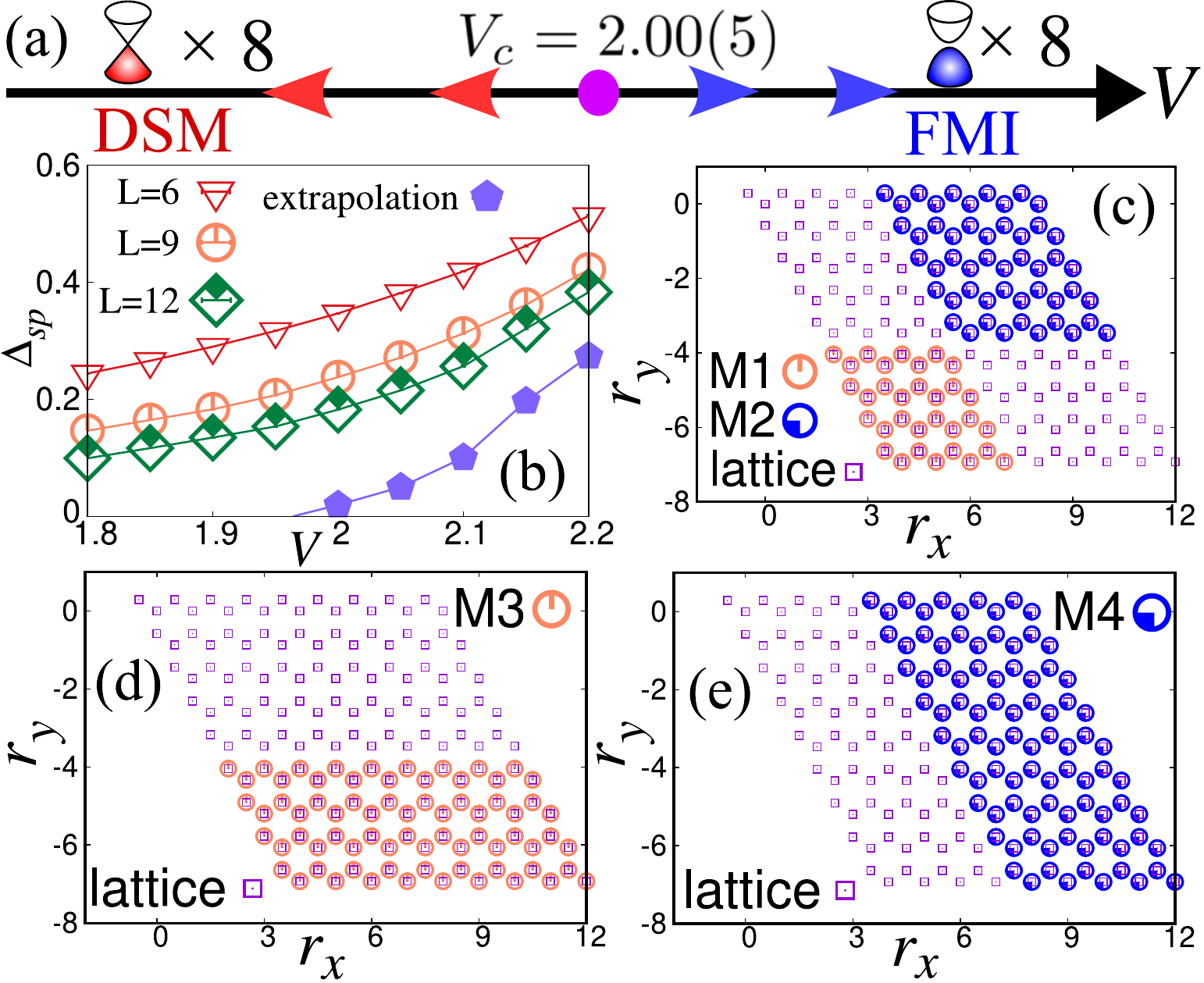}
\caption{\textbf{Phase diagram and entanglement regions.}(a) Ground state phase diagram sketch of our lattice model, featuring the SMG QCP at $V_c$ separating the Dirac semi-metal (DSM) and featureless Mott insulator (FMI) phases. (b) Fermion single particle gap $\Delta_{\text{sp}}$ against interaction strength $V_c$, with the purple curve indicating extrapolated values (refer to SM~\cite{suppl}). (c)(d)(e) Entanglement regions M1, M2, M3, and M4 are highlighted, with M1 and M2 (orange and blue dots in (c)) utilized for determining corner contributions on the disorder operator. Regions M3 (orange dots in (d)) and M4 (blue dots in (e)) represent rectangle stripes wrapped around the $x$ and $y$ directions, free from corner contributions.}
\label{fig:model}
\end{figure}

\begin{figure}[htp!]
\includegraphics[width=\columnwidth]{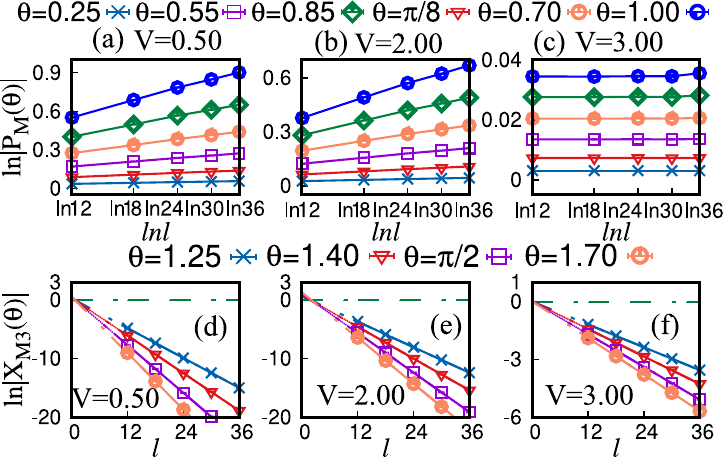}
\caption{\textbf{Disorder operators.} Top panels:  $\ln|P_M(\theta)|$  (Eq.~\eqref{eq:disop_combin}) as function of the perimeter $l=2L$ in the DSM phase (a), at the SMG QCP (b) and in the  FMI phase (c). Data suggests $P_M(\theta)\sim l^{2s(\theta)}$. Bottom panels: The disorder operator $\ln |X_{M3}(\theta)|$ as function of perimeter $l$ in the DSM phase (d), at the SMG QCP (e) and in the FMI phase (f). In the absence of corners, $\ln |X_{M3}(\theta)| \sim -al+\beta(\theta)$.  
}
\label{fig:disop}
\end{figure}

\noindent{\textcolor{blue}{\it Models and numerical settings.}---} We study the following Hamiltonian on the honeycomb lattice 
\begin{align}
\hat{H}= & -t\sum_{\left\langle ij\right\rangle ,\alpha}\left(-1\right)^{\alpha}\left(\hat{c}_{i\alpha}^{\dagger}\hat{c}_{j\alpha}+\hat{c}_{j\alpha}^{\dagger}\hat{c}_{i\alpha}\right)+\nonumber \\
&V\sum_{i}\left(\hat{c}_{i1}^{\dagger}\hat{c}_{i2}\hat{c}_{i3}^{\dagger}\hat{c}_{i4}+\hat{c}_{i4}^{\dagger}\hat{c}_{i3}\hat{c}_{i2}^{\dagger}\hat{c}_{i1}\right)
\label{eq:su4_model}
\end{align}
where the index $\alpha=1,2,3,4$ and $\left\langle\cdots \right\rangle$ reflect the fermion nearest neighbour hopping. At small $V$, the low energy physics of this model is captured by 8 weakly interacting two-component Dirac fermions. The Hamiltonian Eq.~\eqref{eq:su4_model} has a global SU(4) symmetry at half-filling, which manifests after a particle-hole transformation of flavors $\alpha = 2, 4$. This model Eq.~\eqref{eq:su4_model} has been studied with fermion QMC in Ref.~\cite{he2016quantum}, where a SMG QCP was found in the ground state of the model while tuning $V$. Since the phase transition is not driven by spontaneous symmetry breaking, we locate the position of the QCP by the opening of the fermion single particle gap $\Delta_{sp}$, as schematically presented at Fig.\ref{fig:model} (a). As shown in Fig.\ref{fig:model} (b) and in the Supplemental Materials (SM)~\cite{suppl} as well as in Ref.~\cite{he2016quantum}, the single particle gap after extrapolation to the thermodynamic limit opens at the QCP, $V_c=2.00(5)$.   
Ref.~\cite{he2016quantum} further showed that at $V>V_c$, there is no apparent symmetry-breaking of the SU(4) symmetry in the ground state, and the phase diagram is described by a single phase transition from Dirac Semi-metal (DSM) at $V<V_c$ to a featureless Mott insulator (FMI) at $V>V_c$. In this paper, we employ the projector fermion auxiliary field QMC simulations~\cite{Sugiyama86,Sorella89,assaad2008world} with linear system size $L=3,6,9,12,15,18$ and the number of sites $N=2L^2$ for honeycomb lattice (with four flavors of fermions per site), with projection length scaling as $\Theta=2L$ to investigate the ground state properties of the system. We also note that similar SMG transitions between DSM and FMI have been observed in interacting staggered fermion on a cubical space-time lattice with global SU(4) symmetry and there is no spontaneous symmetry breaking, i.e., condensation of fermion bilinear, observed from large-scale QMC simulations~\cite{ayyarMassive2015,ayyarOrigin2016}.
 
\begin{figure}[htp!]
\includegraphics[width=\columnwidth]{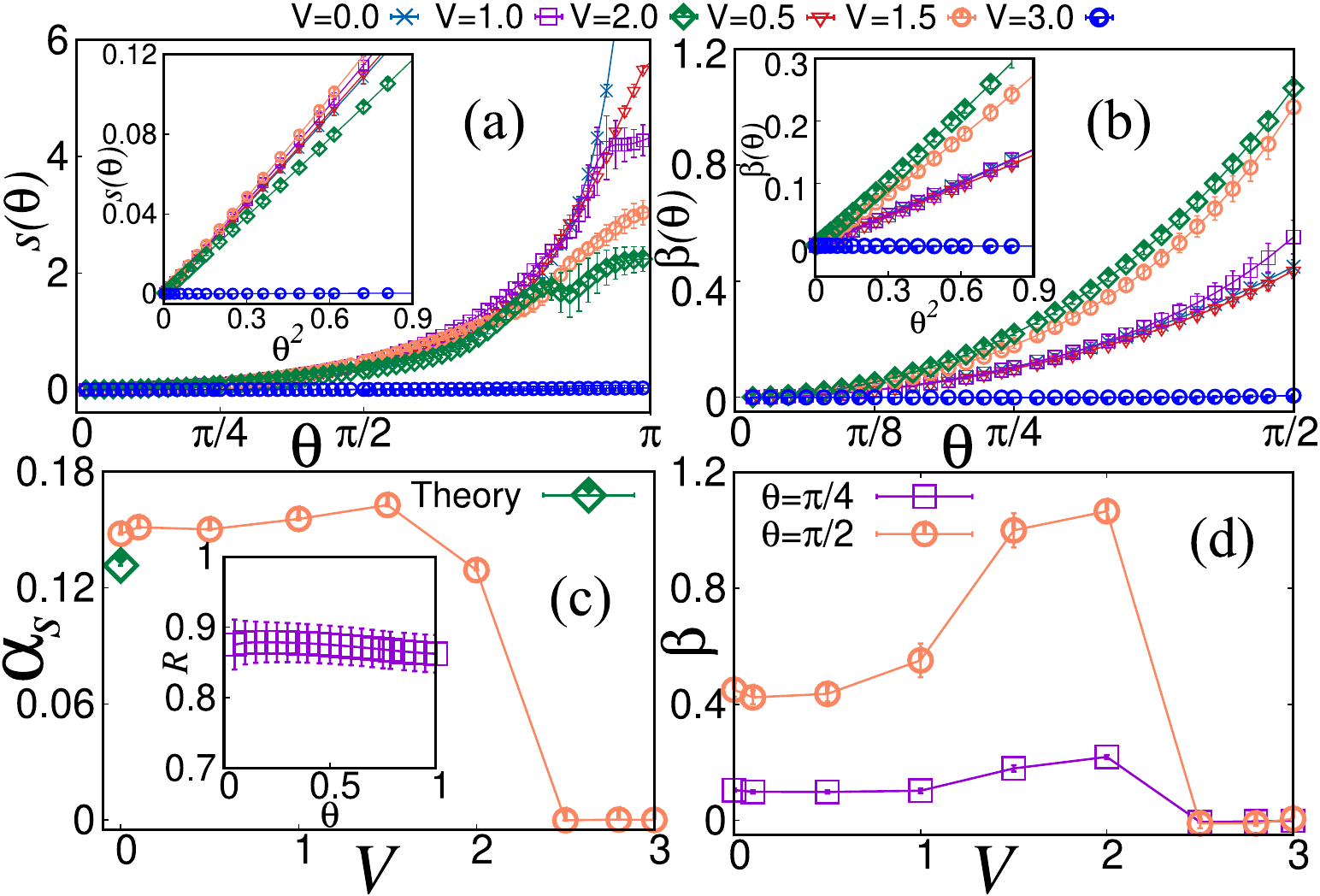}
\caption{\textbf{Logarithmic correction $s(\theta)$ of the disorder operator.} (a) Logarithmic coefficient $s(\theta)$ extracted from $P_M(\theta)$ in top panels of Fig.~\ref{fig:disop}. $s(\theta)$ is positive for all the rotation angle $\theta\in[0,\pi]$. (b)  The constant correction $\beta(\theta)$ extracted from the disorder operator in the bottom panels of Fig.~\ref{fig:disop}, defined in smooth region $M3$ as a function of rotation angle $\theta$. Insets in (a) and (b) show the $x$-axis rescaled to $\theta^2$, reflecting quadratic angle dependence of $s(\theta)$ and $\beta(\theta)$. (c) The quadratic coefficient $\alpha_s$  of $s(\theta)\sim\alpha_s\theta^2$ reduces at the SMG QCP and vanish in the FMI phase. Inset in (c) presents the ratio $R=\frac{s(\theta,V=V_c)}{s(\theta,V=0)}<1$ at small $\theta$, analogous to the Gross-Neveu transition discussed in Ref.~\cite{liu2023fermion}. (d) $\beta(\theta)$ from (b) with two specific $\theta$ values ($\pi/4$ and $\pi/2$) as a function of $V$, a peak of $\beta$ develops at $V_c$ and then drops to zero at $V>V_c$.}
\label{fig:sq}
\end{figure}

\noindent{\textcolor{blue}{\it Numerical probes}---} To probe the intrinsic properties of the SMG QCP, we analyze two quantities in QMC simulation. The first one is the disorder operator defined by the symmetry properties of a system~\cite{fradkin2017disorder,nussinov2009sufficient,nussinov2009symmetry,kadanoff1971determination}. For a $(2+1)d$ theory with at least a U(1) symmetry, one can define a disorder operator as \beqn \hat{X}_M(\theta) = \prod_{i\in M}\exp(\ii \theta \hat{n}_i), \eeqn where $\hat{n}_i$ is the charge density of the U(1) symmetry at site $i$. We are interested in the scaling behavior of the disorder operator. 
If the IR limit of the theory is a $(2+1)d$ CFT, the scaling form of the disorder operator should be dominated by a perimeter law, followed by an additive logarithmic corrections when the region $M$ has corners~\cite{wucorner,wang2021scaling,Estienne_2022}, \beqn \ln\left| X_M(\theta) \right|\sim - a l + s(\theta) \ln l + c, \label{cornerlog} \eeqn where $X_M(\theta) = \langle \hat{X}_M(\theta) \rangle$. In the SM, we present a proof that $s(\theta)$ must be non-negative for all $\theta$, for a class of unitary theories~\cite{suppl}. 

It has been shown that the disorder operator can be conveniently computed in many numerical methods such as QMC and DMRG. The scaling properties of the disorder operator for $(2+1)d$ transverse field Ising, O(2), O(3), topological ordered state and Gross-Neveu transitions have been successfully carried  out~\cite{zhao2021higher,wang2021scaling,wang2022scaling,chen2022topological,jiang2022fermion,liu2023fermion}, and for these theories the log-coefficients of the disorder operators find agreement with unitary CFTs. On the other hand for models that supposedly realize the DQCP (both in spin and fermion realizations)~\cite{wang2022scaling,liu2023fermion}, the scaling of the disorder operator suggests that the transition is not a unitary CFT. 

Since the model Hamiltonian in Eq.~(\ref{eq:su4_model}) has SU(4) flavor symmetry at half-filling~\cite{he2016quantum}, we employ the U(1) charge density operator \beqn \hat{n}_i = \sum_{\alpha = 1}^4 \hat{n}_{i\alpha}-2, \eeqn which is one of the SU(4) symmetry generators.
To extract the subleading logarithmic coefficient $s(\theta)$ in a reliable way, we introduce a new partition strategy to cancel the dominant perimeter law contribution. Following Refs.~\cite{wang2021scaling,kallin2014corner}, we introduce the ratio $P_M(\theta)$ \begin{equation} P_M(\theta)=\left|\frac{X_{M1}(\theta)X_{M2}(\theta)}{X_{M3}(\theta)X_{M4}(\theta)}\right| \label{eq:disop_combin} \end{equation} where $M1$ and $M2$ are the two distinct entanglement regions, shown in Fig.~\ref{fig:model} (c), with the same corner contribution. On the torus geometry, the entanglement regions $M3$ and $M4$ in Fig.~\ref{fig:model} (d) and (e) with shape of $L\times L/2$ are smooth such that the disorder operator is free from corner corrections.  Since the length of the boundary of $M1\cup M2$ is equal to that of $M3\cup M4$, the leading perimeter law scaling is canceled in the quotient $P_M(\theta)$ and we expect $P_M(\theta)\sim l^{2s(\theta)}$. Extracting  $s(\theta)$  from    $P_M(\theta)$  turns  out  to  be  much  more  reliable  than  a  direct  fit of  $X_M(\theta)$ following Eq.~\eqref{cornerlog}.

We also calculate the 2nd order R\'enyi entanglement entropy (EE) $S^{(2)}_M$.
A numerically \textit{cheap}  QMC calculation of the EE~\cite{groverEntanglement2013,Assaad13a} for interacting fermions turns out to be unstable such that a replicated space-time manifold~\cite{calabreseEntanglement2004,Assaad15,Broecker14} has to be used.   This increases the computational complexity which scales as $\beta N^3$ for fermion QMC where $\beta=1/T$ the inverse temperature and $N=2L^2$ for one replica of our honeycomb lattice model. The aforementioned numerical instabilities can be cured by an incremental algorithm developed recently~\cite{demidioUniversal2022,liaoTeaching2023,panComputing2023}. Here, we further employ the improved protocol developed by one of us~\cite{ydliaoControllable2023} to obtain accurate evaluations of the EE. 

\begin{figure}[htp!]
\includegraphics[width=\columnwidth]{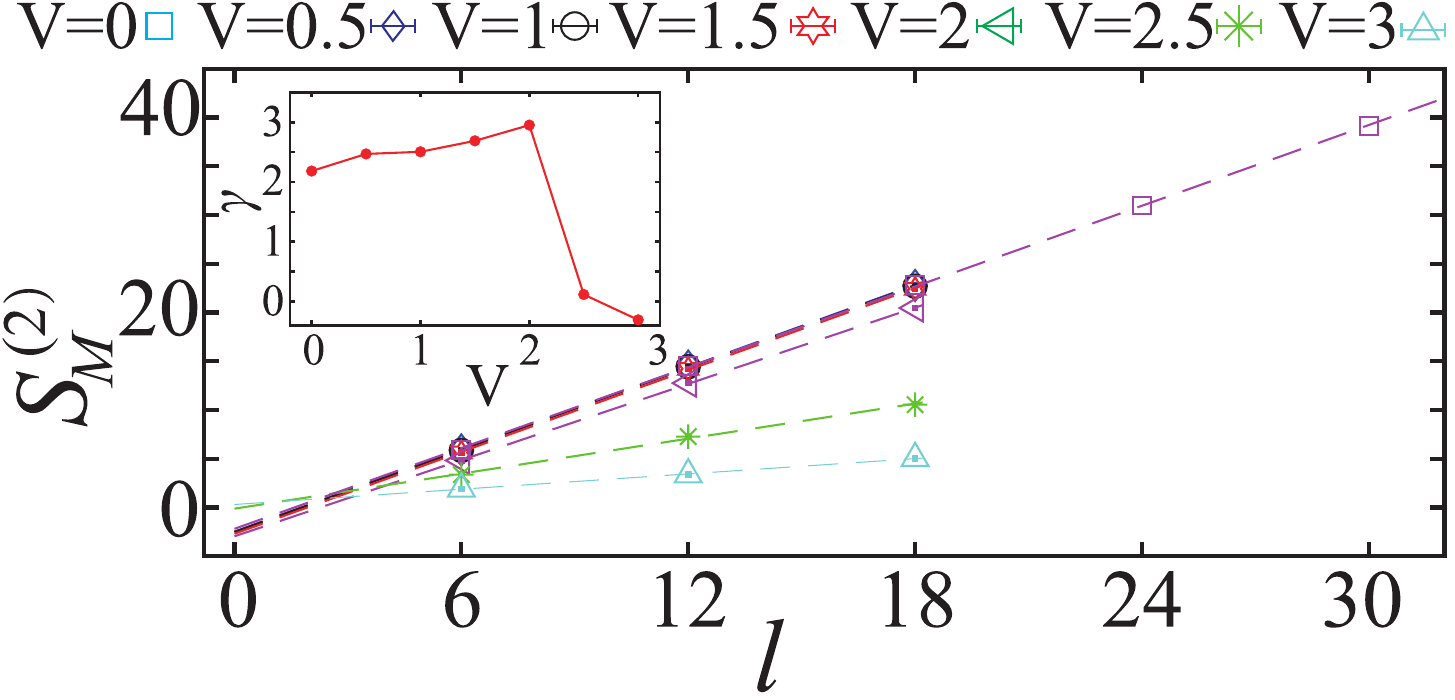}
\caption{\textbf{Entanglement entropy.}  
Finite size scaling relation $S^{(2)}_M = al - \gamma$ of the EE for the bipartition with subregion $L\times L/3$ and $l=2L$. The intercept term $\gamma$ crucially characterizes scaling behavior for different phases. At the SMG QCP $V=V_c$, $\gamma(V_c)$ peaks compared to $\gamma(V<V_c)$ governed by the free Dirac CFT. Entering the FMI phase at $V>V_c$, $\gamma(V>V_c)$ approaches zero, akin to $s(\theta)$ and $\beta(\theta)$ in Fig.~\ref{fig:sq}.}
\label{fig:entropy}
\end{figure}

\noindent{\textcolor{blue}{\it Results of disorder operator.}---} Fig.~\ref{fig:disop} shows our results of $\ln |P_M(\theta)|$ versus $\ln(l)$ for various $\theta$ and $V$ values. According to Eq.~\eqref{eq:disop_combin}, the slope of the curves give rise to the log-coefficient $s(\theta)$ and the obtained results are shown in Fig.~\ref{fig:sq} (a). As was pointed out in previous works~\cite{wucorner,wang2021scaling,liu2023fermion,wangFractional2021}, for small $\theta$, the log-coefficient $s(\theta)$ arising from the corner of the subsystem is proportional to the charge conductivity $\sigma$, which is a universal quantity associated with a $(2+1)d$ CFT~\footnote{Here the conductivity $\sigma$ is in the AC limit, i.e. in the limit $\omega/T \rightarrow \infty$, and $T \rightarrow 0$.}. More precisely, $s(\theta) \sim \alpha_s \theta^2$ for small $\theta$, and the coefficient $\alpha_s$ is proportional to $\sigma$. For example, for a single two-component Dirac fermion in $(2+1)d$ in a region with a sharp corner $\varphi$, it will lead to the following result for the coefficient $\alpha_s$~\cite{wucorner,Estienne_2022}: \beqn 
\alpha_s = \frac{1}{32\pi^2} \left( 1 + (\pi - \varphi)\cot\varphi \right). \eeqn
Both subregions $M_1$ and $M_2$ have two angles with $\varphi = 2\pi/3$, and $\pi/3$, and, in our  case, we  have   $N_f = 8$ flavors of Dirac fermion.  Thus, for  $V < V_c$, in the  DSM phase, the theoretical prediction is $\alpha_s \sim 0.132$, independent of $V$. 
In Fig. \ref{fig:sq} (c), the numerical value of $\alpha_s$ within the DSM phase  stays roughly constant and is larger than the theoretical value indicated by the green dot. The deviation between the numerical results and analytical value and the weak $V$ dependence on $\alpha_s$ inside DSM phase are due to the finite correlation length in QMC simulations. In SM~\cite{suppl}, we provide a mean field analysis to illustrate the finite size effect on computing $\alpha_s$.  We show the value of $\alpha_s$ monotonically converge to the value in the thermodynamic limit by increasing the correlation length at $V=0$.




The SMG transition is essentially a semimetal-insulator transition, intuitively we expect the conductivity at the SMG to be smaller than that in the Dirac semimetal phase. Indeed, the ratio $\alpha_s(V = V_c)/\alpha_s(V = 0)$ is smaller than 1 in our simulation (shown in inset of Fig.~\ref{fig:sq} (c)). It is also the case for the previously investigated Gross-Neveu transition of Dirac fermions, which is also a semimetal-insulator transition~\cite{liu2023fermion,jiang2022fermion}. But we would like to remark that, unlike the EE, there is no general theorem which directly connects the RG flow with the magnitude of $\sigma$. 

It was shown numerically that, at the DQCP $s(\theta)$ can become negative for a broad range of $\theta$~\cite{wang2022scaling,liu2023fermion}, if one tries fitting the data of the disorder operator with the form of Eq.~\ref{cornerlog}, which is in stark contrast against the general conclusion of the non-negativity of $s(\theta)$ for all $\theta$. 
In our model, as shown in Fig.~\ref{fig:sq}(a), $s(\theta)$ indeed remains positive for all $\theta$ and $V \leq V_c$, which is analogous to the situation of the Gross-Neveu transition. When $V > V_c$, $s(\theta)=0$ since the system is gapped and featureless.

If the subregion $M$ has a smooth boundary, i.e. there is no corner, we expect that the disorder operator scales as $\ln \left|X_M(\theta)\right| \sim -al + \beta(\theta) $, with a constant $\beta(\theta)$ which also encodes the information of the CFT.  
The reason for this expectation is that, the conserved charge density that was used to construct the disorder operator is dual to the Wilson loop operator, using the standard dictionary of duality in $(2+1)d$, and the Wilson loop operator can be viewed as a $1d$ defect inserted in the $(2+1)d$ CFT. A quantity similar to $\beta(\theta)$ defined in flat spacetime is referred to as ``defect entropy" in Ref.~\cite{zohardefect}. In Fig.~\ref{fig:sq} (b) we show  $\beta(\theta)$ and the inset exhibits $\beta(\theta)$ versus $\theta^2$. Taking two specific $\theta$ values of $\pi/4$ and $\pi/2$, we observe that $\beta(\theta)$ is finite for $V \leq V_c$, develops a peak close to $V_c$ and a vanishes in the FMI phase, as shown in Fig.~\ref{fig:sq} (d).


\noindent{\textcolor{blue}{\it Results of entanglement entropy.}---} Fig.~\ref{fig:entropy} shows our results for the EE for various $V$ values, 
for corner-free bipartitions of size $L\times L/3$ for $L=3,6,9$ as shown in Fig.~\ref{fig:model} (d) and (e). Since there is no corner contribution in our measurement, we expect the EE to exhibit the scaling form of $S^{(2)}_M = al -\gamma$ with $l=2L$. As shown in Ref.~\cite{chen2017two}, the constant term $\gamma$ of the EE on a torus depends on several global geometric parameters of the entire system and the subregion. In Fig.~\ref{fig:entropy}, we see that the EE is dominated by the perimeter law for all values of $V$. In the inset, we plot the intercepts $\gamma$ as a function of $V$ and observe a maximum at $V = V_c$.
Interestingly, we find that $\beta(\theta)$ from the disorder operator with smooth boundary and $\theta=\pi/2$ behaves similarly to $\gamma$ from EE, where $S_M^{(2)}$ and $X_M$ map onto each other for noninteracting problems~\cite{jiang2022fermion}.

\noindent{\textcolor{blue}{\it Discussions.}---}
In summary, we investigated three quantities that should encode important universal information of the IR behavior of the SMG transition: 

1. For a subregion $M$ with sharp corners, the disorder operator $\hat{X}_M(\theta)$ should scale as $\ln |X_M(\theta)| \sim - a l + s(\theta) \ln l$, where $s(\theta)$ is a universal quantity related to the universal conductivity of the $(2+1)d$ CFT~\cite{Estienne_2022,Giombi_2016,Diab_2016,PhysRevB.88.155109}; and $s(\theta)$ should be non-negative for all $\theta$, for a large class of unitary $(2+1)d$ CFT (see SM~\cite{suppl}). 

2. For a smooth subregion $M$ without corner, the disorder operator $\hat{X}_M(\theta)$ scales as $\ln |X_M(\theta)| \sim - a l + \beta(\theta)$, and $\beta(\theta)$ peaks at the SMG QCP. 

3. For a smooth subregion $M$ without corner, the second R\'enyi EE scales as $S^{(2)}_M \sim  a l - \gamma$, where $\gamma$ is also a universal quantity, which peaks at the SMG QCP. 

The three universal quantities, i.e. $s(\theta)$, $\beta(\theta)$, and $\gamma$ are all nonzrero in the Dirac semimetal phase as well as the SMG QCP, and all vanish inside the FMI phase. In particular, $s(\theta)$ remains positive for all $\theta$ at the SMG QCP. These findings strongly suggest that the SMG transition indeed corresponds to a true $(2+1)d$ unitary CFT, without violating the non-negativity bound as in the case of DQCP.



Various mysteries regarding the SMG transition still remain. For instance, the Dirac semimetal phase with $V < V_c$ should have a large emergent SO(16) symmetry, which manifests when one Dirac fermion is expressed as two Majorana fermions. Although the lattice model breaks $\SO(16)$ symmetry, it is the maximal possible emergent symmetries of the IR fixed points. Whether this $\SO(16)$  symmetry (or its subgroup) can emerge at the SMG transition remains uncertain. Future investigations should examine the disorder operator associated with the generators of SO(16) to ascertain the full emergent symmetry at the SMG. Additionally, verifying the universal corner-log correction of EE at the SMG QCP is pertinent. Due to computational constraints in our current model, extracting this information was unfeasible. However, we aim to explore it using more efficient algorithms in the future.

{\it{Acknowledgment.-}} The authors thank Yin-Chen He and Meng Cheng for very helpful discussions. ZHL acknowledges the Deutsche Forschungsgemeinschaft through the W\"urzburg-Dresden Cluster of Excellence on Complexity and Topology in Quantum Matter -- \textit{ct.qmat} (EXC 2147, Project No.\ 390858490). GPP, MHS, JRZ, WLJ and ZYM acknowledge the support from the Research Grants Council of Hong Kong Special Administrative Region (SAR) of China (Project Nos. 17301721, AoE/P701/20, 17309822, C7037-22GF, 17302223), the ANR/RGC Joint
Research Scheme sponsored by the RGC of Hong Kong
SAR of China and French National Research Agency
(Project No. A\_HKU703/22) and the Beijng PARATERA Tech CO.,Ltd. (URL:https://cloud.paratera.com) for providing HPC resources that have contributed to the research results reported within this paper.
YDL acknowledges support from National Natural Science Foundation of China (Grant No. 12247114). YZY is supported by the National Science Foundation (Grant No.\;DMR-2238360). C. X. is supported by the Simons Investigator program. C.-M.J. is supported by a faculty startup grant at Cornell University. 
FFA acknowledges financial support from the German Research Foundation (DFG) under the grant AS 120/16-1 (Project number 493886309) that is part of the collaborative research project SFB Q-M\&S funded by the Austrian Science Fund (FWF) F 86. 
FFA  and  ZHL  gratefully acknowledge the Gauss Centre for Supercomputing e.V.\ (www.gauss-centre.eu) for funding this project by providing computing time on the GCS Supercomputer SUPERMUC-NG at Leibniz Supercomputing Centre (www.lrz.de),   (project number pn73xu)     
as  well  as  the scientific support and HPC resources provided by the Erlangen National High Performance Computing Center (NHR@FAU) of the Friedrich-Alexander-Universit\"at Erlangen-N\"urnberg (FAU) under the NHR project b133ae. NHR funding is provided by federal and Bavarian state authorities. NHR@FAU hardware is partially funded by the German Research Foundation (DFG) -- 440719683. 
The  calculations  for  the    discorder  
operator  were  carried out  with the  ALF-package \cite{ALF_v2}

\bibliography{SMG.bib}
\bibliographystyle{apsrev4-2}
\startsupplement

\begin{widetext}


\begin{center}
{\bf \uppercase{Supplemental Materials for \\[0.5em]
Disorder Operator and R\'enyi Entanglement Entropy of Symmetric Mass Generation}}
\end{center}

\section{Proof of non-negativity of $s(\theta)$ (for all $\theta$) for a class of theories}

\subsection{a positive-semidefinite matrix $X(\theta)_{ij}$}

Let us use the coordinate system Fig.~\ref{coordinate}. The $2d$ plane in this figure is the actual spatial plane; there is also an out-of-plane direction which is the imaginary-time direction $\tau$.

Just like the proof concerning the Renyi entropy~\cite{casini2012} and the $Z_2$ disorder operator~\cite{wang2022scaling}, our first task is to prove that, for a class of theories, for the $\U(1)$ disorder operator with arbitrary twisting angle $\theta$, the following ``matrix" is positive semidefinite: \beqn X(\theta)_{ij} = \langle \hat{X}(\theta)_{ij} \rangle = \langle \exp\left( \ii \theta \int_{\vect{x} \in M_i \cup \bar{M}_j } d^2x \ \hn(\vect{x}) \right) \rangle. \label{matrix} \eeqn Here $M_i$ and $\bar{M}_i$ are a pair of regions related by reflection $x \rightarrow -x$ (Fig.~\ref{coordinate}).

To prove this statement, we need to show that for any complex vector $\vect{\lambda} = (\lambda_1, \cdots \lambda_N)$, the following is true: \beqn \sum_{i,j} \lambda_i X(\theta)_{ij} \lambda_j^\ast \geq 0. \label{positive}\eeqn The connection between the fact matrix $X(\theta)_{ij}$ being positive semidefinite and Eq.~\ref{positive} is that, one can choose vector $\vect{\lambda}$ that is an eigenvector of $X(\theta)_{ij}$, then Eq.~\ref{positive} is proportional to one of the eigenvalues of $X(\theta)_{ij}$. If Eq.~\ref{positive} is valid for arbitrary $\vect{\lambda}$, then all eigenvalues of $X(\theta)_{ij}$ are non-negative, hence $X(\theta)_{ij}$ is positive semidefinite.

$ $

{\bf Class 1: Complex scalar theories}

Let us first consider the simplest example of theory, i.e. a theory that involves a complex bosonic scalar field $\phi$; but our conclusion in this section can be easily generalized to theories that involve {\it arbitrary numbers} of complex bosons. 
Here we view $\phi$ as a coarse-grained rotor field, and $\hat{n}$ is the rotor number. We consider a class of theories whose Lagrangian density takes the form $\cL = |\partial_\tau \phi|^2 + \cH(\phi)$, where $\cH$ is the potential energy density and it is a functional of $\phi(\vect{x},\tau)$ without temporal derivatives; we also assume $\cH(\phi)$ has spatial reflection symmetry. One necessary condition for the Lagrangian density $\cL$ to take the form above is that, the theory has an exact particle-hole symmetry that takes $\phi(\vect{x}) \ra \phi^\ast(\vect{x})$. One can see the Lagrangian density $\cL(\phi(\vect{x},\tau))$ must be {\it real} at any point of the Euclidean space-time. Needless to say, in the real-time formalism, the Lagrangian density of $\phi$ is real anywhere even without PH symmetry; But in the Euclidean space-time, if the PH symmetry is broken, the following term is allowed: $\phi^\ast \partial_\tau \phi - \phi \partial_\tau \phi^\ast$, and this term is odd under complex conjugation, for bosonic scalar field $\phi$.

Up to a normalization factor, the quantity in Eq.~\ref{positive} is proportional to \beqn \sum_{i,j} \lambda_i X(\theta)_{ij} \lambda_j^\ast &\sim& \int D [\phi_0(y, \tau)] \left( \sum_i \lambda_i Z_{M_i} \right) \left( \sum_j \lambda^\ast_j Z_{\bar{M}_j} \right), \cr\cr Z_{M_i} &=& \int D[\phi(\vect{x}, \tau)]_{\phi(x = 0^-,y,\tau) = \phi_0(y,\tau)} e^{- \int_{x < 0} d^3x  \ \cL^\theta (\phi(\vect{x},\tau))} \cr\cr Z_{\bar{M}_j} &=& \int D[\phi(\vect{x}, \tau)]_{\phi(x = 0^+,y,\tau) = \phi_0(y,\tau)} e^{- \int_{x > 0} d^3x  \ \cL^\theta (\phi(\vect{x},\tau))} \label{Zs} \eeqn Here $\cL^\theta$ is the Lagrangian that includes the action of the disorder operator $\hat{X}(\theta)_{ij}$. The key observation here is that, $\cL^\theta$ {\it is still real} at any point of the Euclidean space-time. To show this, we need to consider the regularized ``connectivity" between the two halves of the space-time $\tau < 0$ and $\tau > 0$. The following regularized term is the most natural choice near the $2d$ plane $\tau = 0$ and $\vect{x}\in M$ (where $M$ could be $M_i$ or $\bar{M}_j$ depending on which partition function to calculate): \beqn \cL^\theta = -e^{\ii \theta} \phi^\ast(\vect{x}, 0^+) \phi(\vect{x}, 0^-) - e^{- \ii \theta} \phi(\vect{x}, 0^+) \phi^\ast(\vect{x}, 0^-) + \cdots \label{eq:clTheta}\eeqn Here, the term explicitly written in Eq. \ref{eq:clTheta} should be viewed as the hopping term along the temporal direction that regularizes the term $|\partial_\tau \phi|^2$ within region $\vect{x}\in M$ and $\tau=0^{\pm}$; For $\vect{x} \notin M$, we set $\theta = 0$. The ``..." part contains terms in $\cL$ that need not be regularized. Obviously, the Lagrangian $\cL^\theta$ is still real for {\it any} configuration of $\phi$ even in the presence of the connectivity terms.


\begin{center}
\begin{figure}
\includegraphics[width=0.4\textwidth]{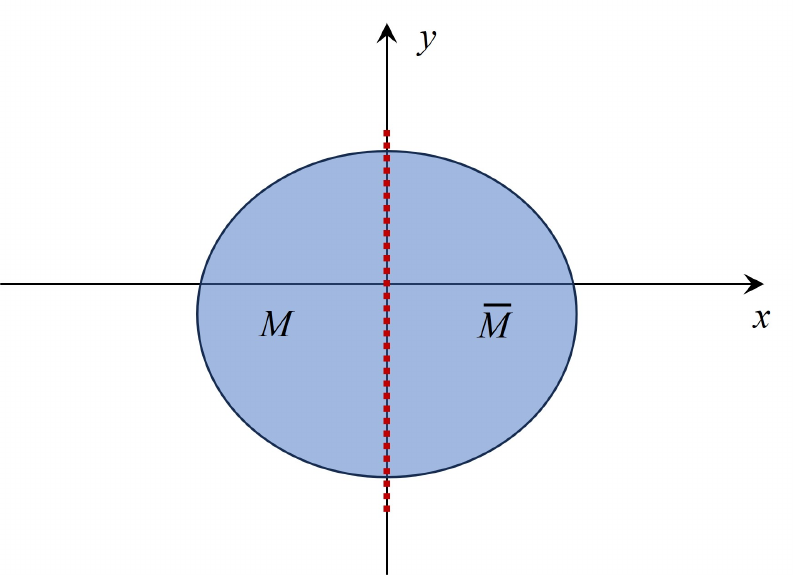}
\caption{The coordinate used in this section of supplementary material. There is an out-of-plane direction that is the imaginary time $\tau$.} \label{coordinate}
\end{figure}
\end{center}

Now it is easy to see that, for {\it any choice} of $\phi_0(y, \tau)$, $Z_{M_i}$ and $Z_{\bar{M}_i}$ are both real and equal to each other, if the system has a particle-hole symmetry ($\phi(\vect{x}) \rightarrow \phi(\vect{x})^*$)  and a reflection symmetry ($(x,y,\tau) \rightarrow (-x,y,\tau)$, $\phi(x,y,\tau) \rightarrow \phi(-x,y,\tau) $). Then for any choice of $\phi_0(y, \tau)$, the quantities in the two parentheses of Eq.~\ref{Zs} are complex-conjugate to each other, for {\it arbitrary choice} of $\lambda_i$, hence the quantity Eq.~\ref{positive} is always non-negative.

$ $

{\bf Class 2: Theories with a Chern-Simons term}

Now we consider theories whose Lagrangian density is {\it not} always real, but we still want the theories to be unitary, hence the imaginary terms of the Lagrangian density will be topological terms. One of the common topological terms is the CS theory. In particular, we consider theories with the following form: \beqn \cL = \cL_0(\phi_\alpha, a_\mu) + \frac{\ii}{4\pi} a \wedge da. \eeqn The term $\cL_0(\phi_\alpha, a_\mu)$ contains minimal coupling between multi-component complex scalar fields $\phi_\alpha$ and the {\it dynamical} gauge field $a_\mu$, as well as other local interactions. We have chosen the level of the CS terms to be $+1$, the reason is that in this way the Hilbert space of the system contains {\it fermions}. And it was proposed that in the infrared $(2+1)d$ Dirac fermions are dual to Chern-Simons matter theories (see for instance Ref.~\cite{seiberg2016duality}). This bosonization approach allows us to conveniently discuss systems with fermions. But the conclusions in this section should be valid for other levels of the CS term.

As a concrete example, let us consider the following choice of $\cL_0(\phi_\alpha, a_\mu)$: \beqn \cL_0(\phi_\alpha, a_\mu) = \sum_{\alpha = 1,2}|(\partial_\tau - \ii a_\tau) \phi_\alpha|^2 + \cH(\phi_\alpha, a_x, a_y) \eeqn With the CS term of $a_\mu$, the Hilbert space of this theory contains two flavors of fermions. We still assume $\cH$ has sufficient discrete symmetries including the PH symmetry, and spatial reflection, which rules out terms such as linear spatial derivative. Also, we make the same assumption as in class 1 that $\cH$ is the potential energy density that does not contain the temporal-derivative of $\phi(\vect{x},\tau)$. Now the disorder operator we consider is \beqn \hat{X}(\theta)_{ij} = \exp\left( \ii \theta \int_{\vect{x} \in M_i \cup \bar{M}_j} d^2x \ (\hat{n}_1 - \hat{n}_2) \right), \eeqn where $\hat{n}_1$ and $\hat{n}_2$ are the rotor numbers of $\phi_1$ and $\phi_2$ respectively. The $\U(1)$ rotation we perform is a {\it global symmetry}, rather than part of the gauge $\U(1)$ group. It is important to note that, $\cL^{\theta}_0$ is real everywhere for any configurations of $\phi$, $a_\mu$ and $\theta$, the imaginary part of the Lagrangian only arises from the CS term. We need two flavors of the bosonic fields to have an internal global U(1) symmetry that is independent from the gauge group, but from this point we are going to hide the index $\alpha$ for the sake of conciseness.

For convenience, let us take the {\it Landau gauge} $a_x(\vect{x},\tau) = 0$. In this gauge, the CS term only involves derivatives of $x$: \beqn \frac{\ii}{4\pi} a \wedge da = \frac{\ii}{4\pi} ( a_y \partial_x a_\tau  - a_\tau \partial_x a_y). \eeqn In the path integral we impose the following boundary conditions\beqn x < 0, \ \ \phi(x = 0^-,y,
\tau) = \phi_0(y,\tau), \ \ \vect{a}(x = 0^-,y,\tau) = \vect{a}_0(y,\tau), \cr\cr x > 0, \ \ \phi(x = 0^+,y,\tau) = \phi_0(y,\tau), \ \ \vect{a}(x = 0^+,y,\tau) = \vect{a}_0(y,\tau). \eeqn Notice that here $\vect{a}_0$ is the boundary conditions of the two nonzero component vector $\vect{a} = (a_\tau, a_y)$, not to be confused with the temporal component of the gauge field.

Now the quantity Eq.~\ref{positive} is evaluated as \beqn \sum_{i,j} \lambda_i X(\theta)_{ij} \lambda_j^\ast &\sim& \int D [\phi_0(y, \tau)] D [\vect{a}_0(y, \tau)] \left( \sum_i \lambda_i Z_{M_i} \right) \left( \sum_j \lambda^\ast_j Z_{\bar{M}_j} \right), \cr\cr Z_{M_i} &=& \int D[\phi(\vect{x}, \tau)] D[\vect{a} (\vect{x}, \tau)] e^{- \int_{x < 0} d^3x \ \cL^\theta (\phi, \ \vect{a})}, \cr\cr Z_{\bar{M}_j} &=& \int D[\phi(\vect{x}, \tau)] D[\vect{a} (\vect{x}, \tau)] e^{- \int_{x > 0} d^3x  \ \cL^\theta (\phi, \ \vect{a})}. \eeqn One can show that, for any configuration $\phi(x, y, \tau)$ and $\vect{a}(x, y,\tau)$, there exists another configuration \beqn \phi'(x, y, \tau) = \phi(-x, y, \tau), \ \ \vect{a}'(x, y, \tau) = \vect{a}(- x, y, \tau), \eeqn which makes \beqn \cL^\theta(\phi'(- x, y, \tau), \vect{a}'(- x, y, \tau)) = \left( \cL^\theta( \phi(x, y, \tau), \vect{a}(x, y, \tau)) \right)^\ast. \eeqn Hence we can conclude that \beqn Z_{\bar{M}_i} = \left( Z_{M_i} \right)^\ast, \eeqn which leads to the conclusion that Eq.~\ref{positive} is still valid. Here we have used two assumptions: ({\it i.}) $\cL^{\theta}$ is real everywhere except for the purely imaginary CS term; and ({\it ii.}) $a_x$ is zero due to the Landau gauge we chose.

We can also prove the same result without choosing a Landau gauge on the entire space-time: for any configuration $\phi(x, y, \tau)$ and $a_\mu(x, y,\tau)$, there exists another configuration \beqn \phi'(x, y, \tau) = \phi(-x, y, \tau), \ \ a'_{y,\tau}(x, y, \tau) = a_{y,\tau}(- x, y, \tau), \ \ a'_{x}(x, y, \tau) = - a_{x}(-x, y, \tau), \eeqn which still guarantees that \beqn \cL^\theta(\phi'(- x, y, \tau), a'_\mu(- x, y, \tau)) = \left( \cL^\theta( \phi(x, y, \tau), a_\mu(x, y, \tau)) \right)^\ast. \eeqn This equation simply states that the theory has a $PT$ invariance. Notice that a CS term of a gauge field breaks $P$ and $T$ separately, but it keeps $PT$. The time-reversal symmetry acts a complex conjugation in the Euclidean space-time. This would still make Eq.~\ref{positive} valid, and now we only need to require $a_x(0, y, \tau) = 0$. This is a much weaker ``gauge condition" that we need to choose.

\begin{center}
\begin{figure}
\includegraphics[width=0.7\textwidth]{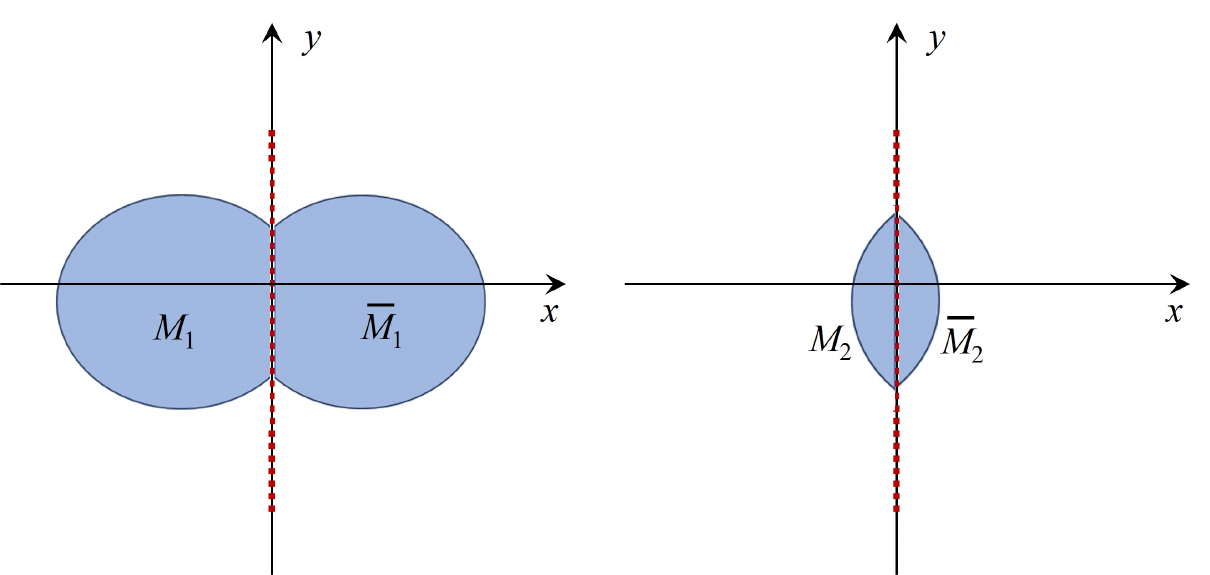}
\caption{We choose regions $M_1$, $M_2$, $\bar{M}_1$, $\bar{M}_2$ for the proof of the non-negativity of $s(\theta)$; $M_1 \cup \bar{M}_2$ and $M_2 \cup \bar{M}_1$ both form a complete oval shape without corner. } \label{shapes}
\end{figure}
\end{center}

\subsection{The corner-log contribution}

After establishing Eq.~\ref{positive} for arbitrary $\theta$ for the classes of theories under consideration, the rest of the discussion follows directly from the appendix of Ref.~\cite{wang2022scaling}. But for the purpose of completeness and self-contained presentation, we finish the whole proof here.

Let us take $i,j = 1,2$ in Eq.~\ref{positive}, and the regions $M_1$, $M_2$, $\bar{M}_1$ and $\bar{M}_2$ are given in Fig.~\ref{shapes}. The inequality Eq.~\ref{positive} implies that \beqn X(\theta)_{11} X(\theta)_{22} - X(\theta)_{12}X(\theta)_{21} \geq 0. \label{positive2} \eeqn Suppose we write $X(\theta)_{ij} = \exp( - S(\theta)_{ij})$, Eq.~\ref{positive2} implies that \beqn S(\theta)_{M_1 \cup \bar{M}_2} + S(\theta)_{M_2 \cup \bar{M}_1} \geq S(\theta)_{M_1 \cup \bar{M}_1} + S(\theta)_{M_2 \cup \bar{M}_2}. \label{inequality}\eeqn Notice that $M_1 \cup \bar{M}_2$ and $M_2 \cup \bar{M}_1$ both form a full oval region without any corner, only regions $M_1 \cup \bar{M}_1$ and $M_2 \cup \bar{M}_2$ have sharp corners and hence potentially contribute to a logarithmic term. More specifically, the region $M_1 \cup \bar{M}_1$ has two corners each with angle $\varphi$, and region $M_2 \cup \bar{M}_2$ has two corners with angle $2\pi - \varphi$. If we write \beqn S(\theta)_M = a l_M - s(\theta)\ln l + \cdots \eeqn and plug this in Eq.~\ref{inequality}, it is easy to see that the first term of perimeter law cancels out in Eq.~\ref{inequality}, and we are left with \beqn s(\theta, \varphi) \ln l + s(\theta, 2\pi - \varphi) \ln l \geq 0. \eeqn

Another relation we need to use is $s(\theta, \varphi) = s(\theta, 2\pi - \varphi)$, this is because ({\it i.}) for a quantum system with an exact $\U(1)$ symmetry, the $X(\theta)_{M}$ defined on a region $M$ with angle $\varphi$ must equal to the $X(- \theta)_{\cA - M}$ defined on region $\cA - M$, where $\cA$ represents the entire $2d$ area $\cA$, and region $\cA - M$ must have angle $2\pi - \varphi$; and ({\it ii.}) $s(\theta) = s(-\theta)$ for a system with particle-hole symmetry. Then we complete the proof that $s(\theta,\varphi) \geq 0$.

Let us review a few key points of the proof.

1. A {\it particle-hole symmetry} of the associated $\U(1)$ global symmetry is a key assumption made in this section. This PH symmetry prohibits some unfavorable terms in the Lagrangian.

2. A direct treatment of the Dirac fermion Lagrangian would be tedious. 
As an alternative approach, we treat fermionic systems in this section as bosons coupled with a CS field. 

3. We do not start with a Lorentz-invariant field theory in our discussion, as the corner-log contribution may arise in theories without any emergent Lorentz invariance, like the logarithmic correction from corners to entanglement entropy of $(2+1)d$ $z = 2$ theories discussed in Ref.~\cite{fradkinmoore}.

\section{Benchmark with Free Dirac systems on honeycomb and $\pi$-flux square lattices}

In this section, we explain how to obtain the R\'enyi entanglement entropy from the fermion Green's function for  lattice models and discuss the impact on log-coefficient caused by different choices of the boundary for the entanglement region for free Dirac  systems on honeycomb and $\pi$-flux square lattices. These results provide a benchmark 
for  our computations  with respect to the exact solvable limit. In \ref{sec:III}, we  present  supplemental  QMC data for the SMG model, including the finite size scaling of the single-particle gap that  used  to pin down  the position of the SMG QCP, and the analysis of the disordered operator in the  free Dirac fermion limit.

For $V=0$ case, when the Hamiltonian on the honeycomb lattice   read
$
\hat{H}_0=  -t\sum_{\left\langle ij\right\rangle ,\alpha}\left(-1\right)^{\alpha}\left(\hat{c}_{i\alpha}^{\dagger}\hat{c}_{j\alpha}+\hat{c}_{j\alpha}^{\dagger}\hat{c}_{i\alpha}\right)$, where the index $\alpha=1,2,3,4$, we can calculate the R\'enyi  entanglement entropy~\cite{Peschel03,casini2009entanglement,panComputing2023,groverEntanglement2013} as:
\begin{equation}
S^{(2)}_M=-\log(\det(G_M^2+(I-G_M)^2)) = a l-\gamma
\end{equation}
 with the length of the entangling boundary $l$.  Here,  $M$ is the entanglement region and $G_M$ is the  equal time Green  function  matrix  restricted to   lattice   sites in  $M$. As a benchmark, the entanglement region we choose is $L\times L/3$ without corner  such  that  $l=2L$.  The boundary of the entangled region is parallel to the primitive vectors  and the constant $\gamma$ is related to the number of Dirac cone. For  our  SU(4)  symmetric  model  we  have  8  two-component Dirac cones. 
 

\begin{figure}[htp!]
\includegraphics[width=0.6\columnwidth]{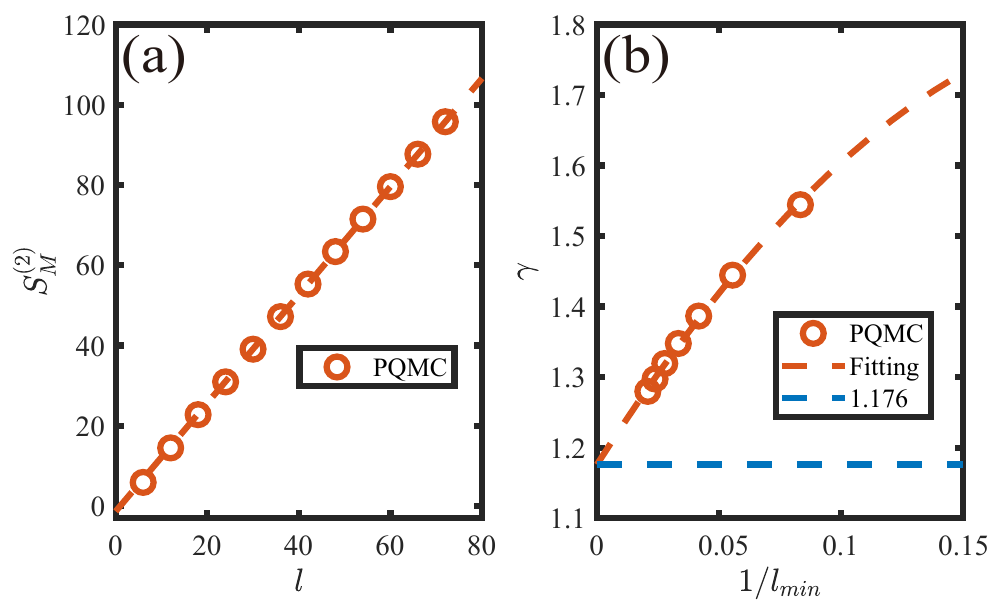}
\caption{\textbf{Benchmark of free fermion part with PBC} (a)  R\'enyi entanglement entropy calculated in PQMC frame for $V=0$ and  the entanglement entropy region is $L\times L/3$, with $l=2L$ in our setting .  (b) Results of fitting and extrapolation. The dark red points represent the intercept $\gamma$ obtained by a linear fit from the point starting at $l_{min}$ while the light blue line is its extrapolation $\gamma \approx 1.176$ }
\label{fig:s1}
\end{figure}

We add a tiny twist (a small shift in momentum space) to lift the degeneracy at Dirac point, which is implemented as a Peierls factor as:
\begin{equation}
H(L)=-t \sum_{\langle\vec{i}, \vec{j}\rangle ,\alpha} e^{\left(i 2 \pi / \Phi_0\right) \int_{\vec{i}}^{\vec{j}} \vec{A}_L(\vec{l}) \cdot d \vec{l}} \hat{c}_{i,\alpha}^{\dagger} \hat{c}_{j,\alpha}+\text { H.c. }
\end{equation}
 We choose the appropriate  $ \vec{A} = \Phi  \vec{a}_1 /L  $ with  $\Phi = 0.00001 \Phi_0$,    with  $\vec{a}_1$   a  lattice  vector    and   periodic  boundary conditions  for  the  fermion operators. 

In Fig.~\ref{fig:s1}(a), we show  $S^{(2)}_M$ as calculated by the Green's function obtained in projector quantum Monte Carlo (QMC) with $V=0$, setting the projection length scaling to $\Theta=2L$.  As shown in Fig.~\ref{fig:s1}(b), we select the data points starting from $l_{min}$ for fitting, and the  extrapolation in  $l_{min}$  provides  an  estimate 
in the  thermodynamic limit,  $\gamma  \approx 1.176$.    We note that its value is different from that $\sim 2$ shown in the inset Fig~\ref{fig:entropy}, which is only obtained from very small size $9\times 9$ with expected finite size effect.

\begin{figure}[htp!]
 	\begin{minipage}[htbp]{0.3\columnwidth}
		\centering
		\includegraphics[width=0.8\columnwidth]{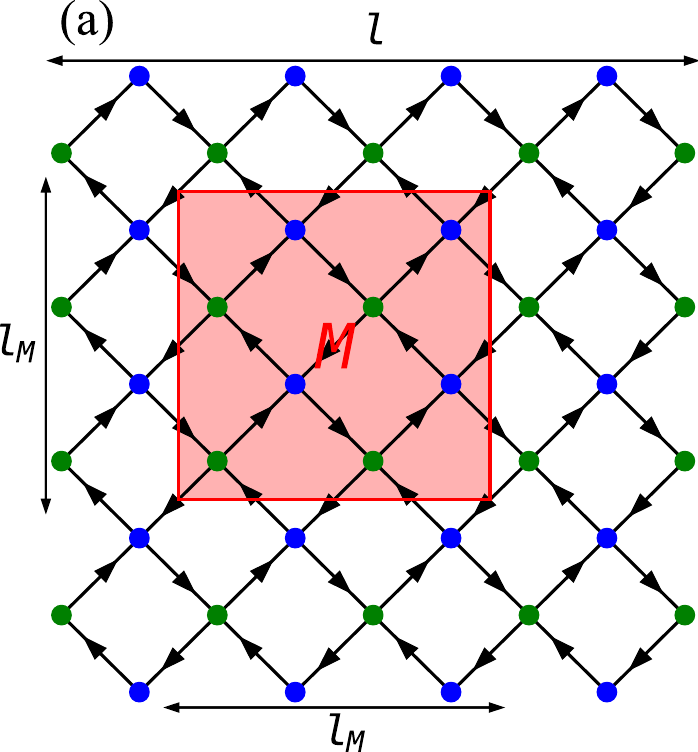}
	\end{minipage}
	\begin{minipage}[htbp]{0.3\columnwidth}
		\centering
		\includegraphics[width=0.8\columnwidth]{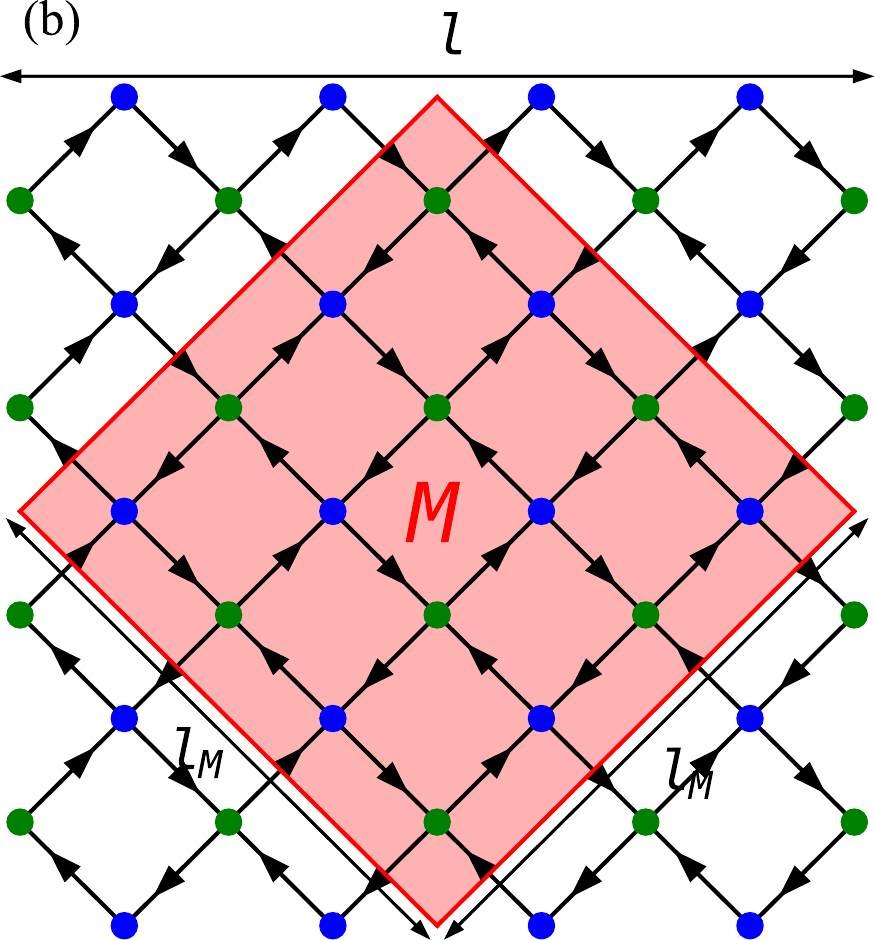}
	\end{minipage}
 	\begin{minipage}[htbp]{0.3\columnwidth}
		\centering
		\includegraphics[width=0.8\columnwidth]{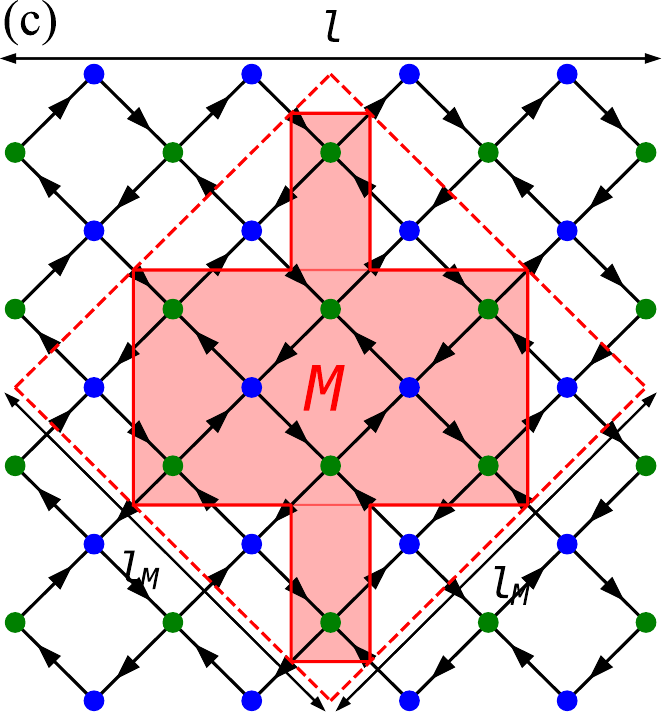}
	\end{minipage}

        \vspace{1cm}
	\begin{minipage}[htbp]{0.3\columnwidth}
		\centering
		\includegraphics[width=\columnwidth]{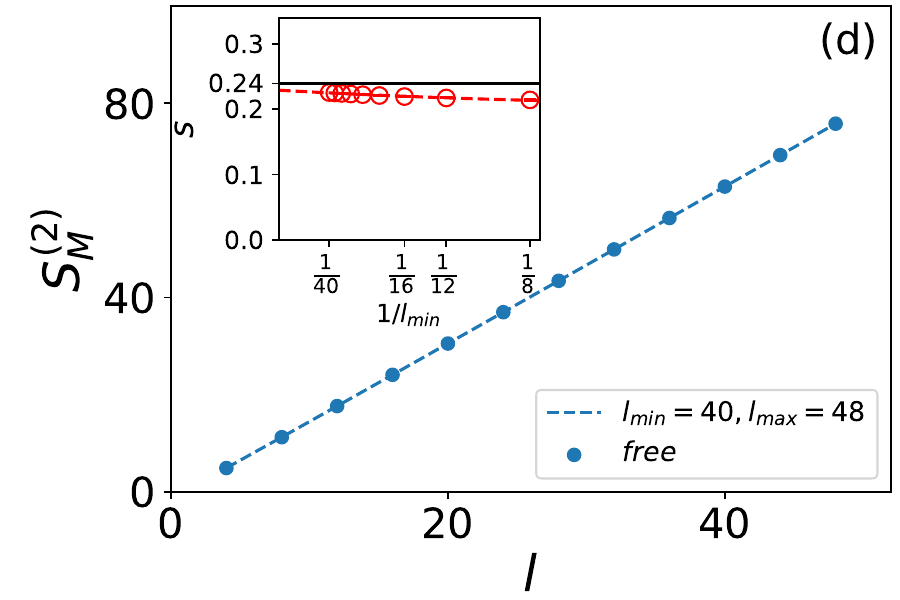}
	\end{minipage}
	\begin{minipage}[htbp]{0.3\columnwidth}
		\centering
		\includegraphics[width=\columnwidth]{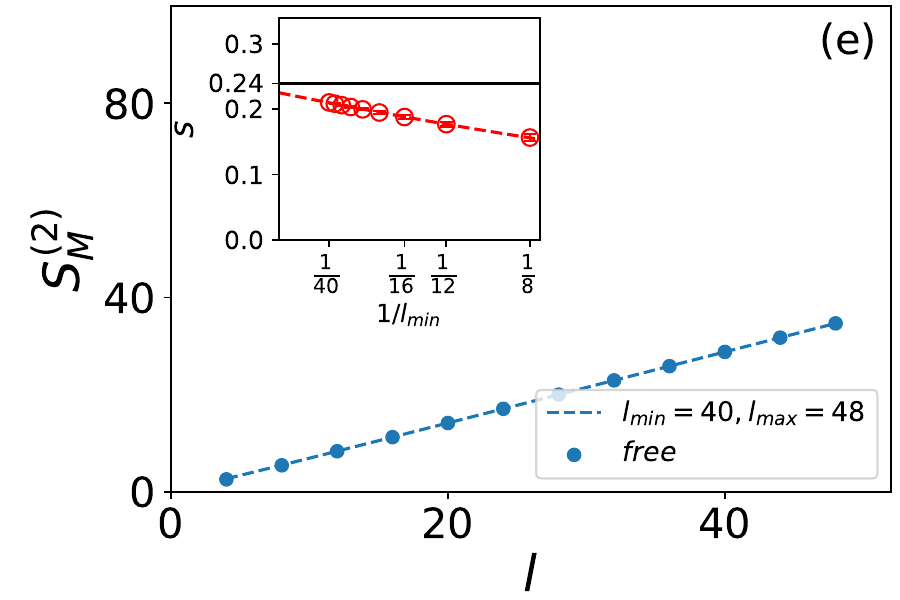}
	\end{minipage}
	\begin{minipage}[htbp]{0.3\columnwidth}
		\centering
		\includegraphics[width=\columnwidth]{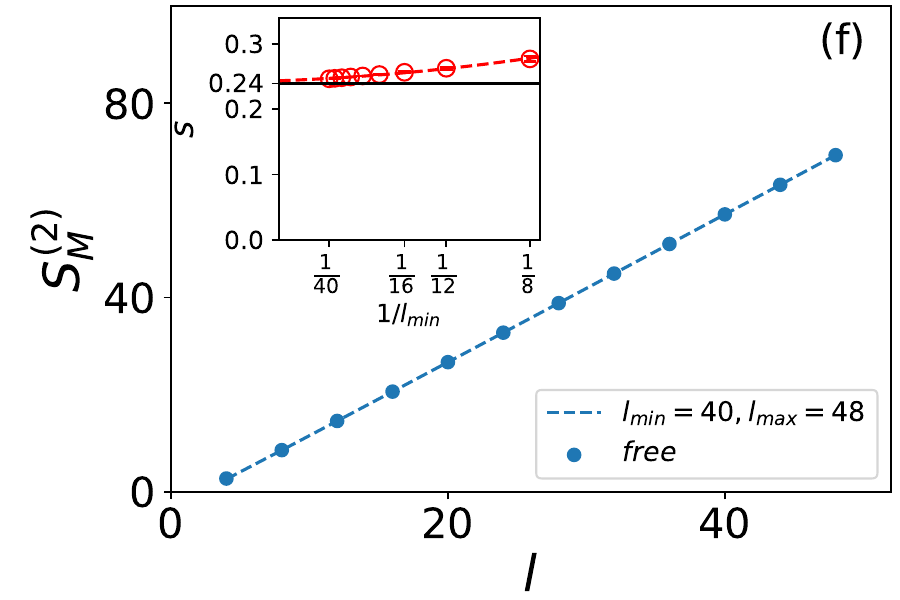}
	\end{minipage}
        \caption{\textbf{Benchmark of free fermion part with various shape of entanglement region.} (a-c) The sketch map for three types of boundary considered in this paper.  (a) The boundary of entangled region is parallel to the system boundary, which contains integer number of unit cells. (b) 45 degree included angle for the entangle region, where the sublattice is cut in one unit cell. (c) According to (b), one form zig-zag shape for the 45 degree boundary. That is to say, for the site near the boundary, only one sort of the sublattice is included. (d-e) The entanglement entropy and extrapolation results for three types of boundary. All of them have extrapolation results closed to 0.24.}
	\label{fig:s2}
\end{figure}

We then explore the impact caused by distinct choice of the boundary for the entanglement region. Here we focus on the $\pi$-flux square lattice at its non-interacting limit as shown in Fig.~\ref{fig:s2} (a). The model and its disorder operator and R\'enyi EE at its non-interacting limit and the Gross-Neveu chiral Ising QCP with interaction have been thorougly discussed in the Sec. III of the Supplemental Materials of Ref.~\cite{liu2023fermion} and in Ref.~\cite{liaoTeaching2023}, respectively. The reason of choosing $\pi$-flux quare lattice over the honeycomb lattice is that it is much easier to generate the entanglement region $M$ with different boundaries as denoted in Fig.~\ref{fig:s2} (a), (b) and (c).

We focus on the log-correction, determined by the corner contribution. We still follow previous setting, choosing a twist field 0.001 in  both directions at  zero temperature. We change the system size $L$ and fix $L_M = \frac{L}{2}$. Three types of boundary condition are considered, shown in Fig.~\ref{fig:s2} (a), (b) and (c). One 90 degree corner of $M$ contributes
to a universal coefficient of 0.01496 to the log term in the
Dirac CFT~\cite{helmes2016universal}. Since in our model, rectangle region $M$ has 4 corners and we have two spin flavors and two Dirac cones in the Brillouin zone. The final contribution to the log coefficient is
$s = 16 \times 0.01496 \approx 0.23936$. The differences of the three types of boundaries are (a) The boundary of entanglement region is parallel to the system boundary. (b) The boundaries has an 45 included angle. (c) Based on (b), the boundary has zig-zag form instead a smooth edge. We calculate the R\'enyi entanglement entropy for three cases, and extrapolate to the infinite size at the thermodynamic limit. We find convergent value to 0.24, as shown in Fig.~\ref{fig:s2} (d), (e) and (f), consistent with the analytic results.

\section{QMC measurements}
\label{sec:III}

\subsection{Time displaced Green's function}
To locate the quantum critical point of our model, we compute the dynamic single particle Green's function in the QMC simulation, which is defined by $G(k,\tau)=\sum_{\alpha}\left\langle\hat{c}_{k,\alpha,\tau}\hat{c}^{\dagger}_{k,\alpha,0}\right\rangle$ . In the ground state, the Green's function scale as $G(k,\tau)\propto \exp(-\tau\Delta_{sp}(k))$ in the long time limit $\tau\rightarrow \infty$, where $\Delta_{sp}(k)$ is the single particle gap at momentum $k$. In Fig. \ref{fig:s2}(a), we observe linear imaginary time dependence of Green's function under logarithmic scale.   The finite size single particle gap $\Delta_{sp}(k,1/L)$ are extracted from the slope of these linear data lines. 

To obtain the single particle gap in thermodynamic limit, we use the quadratic polynomial functional fitting to extrapolate the data  $\Delta_{sp}(k,1/L)$  to $1/L=0$ limit. The detail of the numerical extrapolation are presented in Fig. \ref{fig:s2}(b). As we discussed in the main text, the emergence of finite single particle gap in thermodynamic limit provide us the position of the QCP.

\begin{figure}[h!]
\includegraphics[width=0.8\columnwidth]{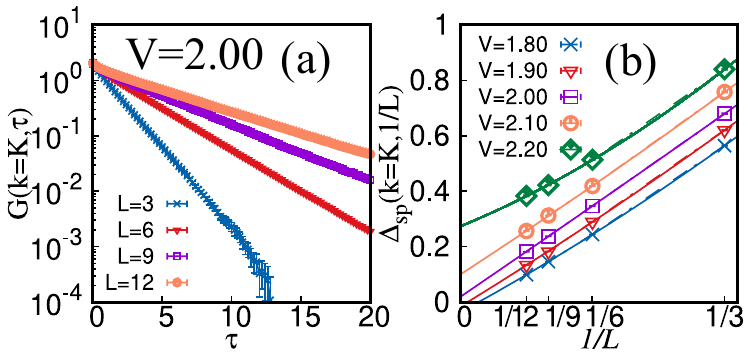}
\caption{\textbf{Time displaced single particle Green's function} (a) Time displaced single particle Green's function at Dirac point $k=\boldsymbol{K}$ as function of imaginary time $\tau$. (b) Extrapolation of single particle gap at Dirac point $\Delta_{sp}(k=\boldsymbol{K})$ with inverse of system size $1/L$.}
\label{fig:s3}
\end{figure}

\subsection{Disorder operator in free Dirac fermion}

In this section, we discuss the behavior of disorder operator in the mean field type lattice model. As a starting point, we consider the following hamiltonian residing on the honeycomb lattice 

\begin{align}
H_{\text{MF}}= & -t\sum_{\left\langle ij\right\rangle \lambda}\left(\hat{c}_{i\lambda}^{\dagger}\hat{c}_{j\lambda}+h.c.\right)+m\sum_{i\lambda}(-1)^{\boldsymbol{i}}\hat{c}_{i\lambda}^{\dagger}\hat{c}_{i\lambda}\label{eq:ham_mft}
\end{align}
where the spin index $\lambda=\{\uparrow,\, \downarrow\}$. The bracket $\left\langle\cdots\right\rangle$ denotes the nearest neighbour hopping on the  honeycomb lattice. The second term on the right hand side of Eq.(\ref{eq:ham_mft}) denotes the charge density mass term and value of the mass is controlled by $m$. The mean field hamiltonian only contains fermion bilinear terms hence we can solve the system on large system sizes. 

At $m=0$, the low-energy effective hamiltonian of the lattice model is described by CFT fixed point. For two dimensional free Dirac system, the conductivity (which is proportional to the current central charge) $\sigma$ can be computed analytically. As we discussed in the main text, one can extract $\sigma$ by computing  the  quotient $P_M(\theta) $.  We use the disorder operator in the charge channel $X_c(\theta)=\left\langle\prod_{i\in M}\exp(i\theta\sum_{\lambda}\hat{c}^{\dagger}_{i\lambda}\hat{c}_{i\lambda}\right\rangle$ and the entanglement  regions are the same as the main text Fig.\ref{fig:model}. 

In Fig. \ref{fig:s4}  (a), we present the results of  the  quantity $P_M(\theta)$ as function of perimeter $l$ for free Dirac fermion systems.   With a  logarithmic scale, the logarithmic correction
correspond  to  the slope of  $P_M(\theta)$.  In the data analysis, we pick up the data using in linear regression from the fitting window $[l_{min},l_{max}]$. In Fig. \ref{fig:s4}  (b), we show
how  the quadratic coefficient $\alpha_s$   varies  as a  function of  the fitting window. We add the green line to indicate the theoretical value  $\alpha_s\approx 0.066$\cite{liu2023fermion}. The value $\alpha_s$ monotonically converges to the  theoretical value as function of $1/l_{min}$ and $1/l_{max}$. For the maximum perimeter $l_{max}=2L=84$, the fitting window with $l_{min}>18$ provides an estimate of $\alpha_s$ which is less than 10 percent different  than  the theoretical value.

We are also interest in the finite size effect of the  logarithmic correction $s(\theta)$. In the thermodynamic limit, we know that  the  logarithmic correction $s(\theta)$  vanishes  in the presence of a finite  mass  $m$. However, on  finite size lattices, the correlation length   may  exceed  the  lattice  size  such that  $s(\theta)$ is not exactly zero. In Fig. \ref{fig:s4}  (b), we fix the fitting window $[l_{min}=12,l_{max}=78]$ and shown the decay of $s(\theta)$ as function of mass gap $m$. The tuning of mass gap can be considered as a way to control the correlation length of the system.  Our mean field calculation suggest exponentially decay of $s(\theta)$  as  the  mass  gap  increases. 

\begin{figure}[h!]
\includegraphics[width=\columnwidth]{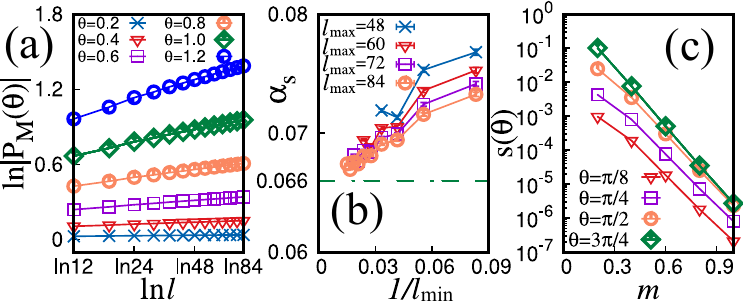}
\caption{\textbf{Disorder operator of free Dirac fermion} (a) Quantity $P_M(\theta)$ for free Dirac fermion on honeycomb lattice. (b) Convergence behavior of quadratic coefficient $\alpha_s$ as function of fitting range $[l_{min},l_{max}]$ . (c) The decay of the logarithmic correction $s(\theta)$ as function of CDW mass $m$, where $s(\theta)$ are obtained with fitting range $[l_{min}=12,l_{max}=78]$.}
\label{fig:s4}
\end{figure}

\end{widetext}

\end{document}